 \documentstyle[12pt]{article}
\newcommand{\be}{\begin{equation}}
\newcommand{\ee}{\end{equation}}
\newcommand{\ptp}{P\"oschl-Teller }

\begin{document}


 \begin{flushright}
    CALT-68-1949, USC-94-015\\
          hep-th/9409097 \\
          September 1994
    \end{flushright}
 \begin{center}
 { \Large \bf On the non-relativistic limit of the quantum sine-Gordon model
 with  integrable boundary condition.}
  \end{center}

 \vspace{4mm}

\centerline{A. ~Kapustin$^\dagger$, S. ~Skorik$^\ddagger$ }
 \vspace{1mm}
\centerline{$^\dagger$ \it Physics Department, California Institute
 of Technology, Pasadena, CA 91125.}
\centerline{$^\ddagger$ \it Physics Department,
University of Southern California, Los Angeles, CA 90089-0484.}

 \begin{abstract}
 We show that the the generalized Calogero-Moser model with boundary potential
of the P\"oschl-Teller type describes the non-relativistic
limit of the quantum sine-Gordon model on a half-line with Dirichlet
boundary condition.

 \end{abstract}


 \vspace{2mm}

In this Letter we consider the sine-Gordon model on a half-line,
\be
{\cal{L}}_{SG}={1\over 2}\int_0^{+\infty}
\left[\left(\partial_t\varphi\right)^2 -
\left(\partial_x\varphi\right)^2 +
{m_0^2\over\beta^2}\cos\beta\varphi\right]dx ~+~ M\cos
{\beta\over 2}(\varphi(x=0)-\varphi_0), \label{lagr}
\ee
with the fixed value of field at the boundary: $\varphi(x=0,t)=
\varphi_0$, or $M=\infty$ in (\ref{lagr}). Such a model was discussed
in \cite{GZ}, where  its quantum integrability and  exact S-matrix were
conjectured. The boundary scattering matrix is diagonal and, according to
 \cite{GZ}, the reflection amplitude of the soliton $P_+$ (resp. $P_-$ for
anti-soliton) reads:
\be
P_{\pm}(\theta)=\cos(\xi\pm \lambda u) R(u,\xi) = \cos(\xi\pm
\lambda u)R_0(u)R_1(u,\xi), \label{reflamp}
\ee
where $\theta=iu$ is the rapidity, $\xi={4\pi\over\beta}\varphi_0$
and $\lambda = {8\pi\over\beta^2}-1$;
 \begin{eqnarray}
R_{0}(u)&=&{\Gamma\left(1-{2\lambda u\over\pi}\right)
\Gamma\left(\lambda+{2\lambda u \over\pi}\right)
\over
\Gamma\left(1+{2\lambda u \over\pi}\right)
\Gamma\left(\lambda-{2\lambda u\over\pi}\right)}
\prod_{k=1}^{\infty}{{\Gamma\left(4\lambda k - {2\lambda u\over\pi}\right)
\over
\Gamma\left(4\lambda k+{2\lambda u \over\pi}\right)}}
\times \nonumber \\
&\times &{\Gamma\left(1 + 4\lambda k -{2\lambda u\over\pi}\right)
\Gamma\left(\lambda(4k+1)+{2\lambda u \over\pi}\right)
\Gamma\left(1+\lambda(4k-1)+{2\lambda u \over\pi}\right)
\over
 \Gamma\left(1+4\lambda k+{2\lambda u \over\pi}\right)
\Gamma\left(\lambda(4k+1) -{2\lambda u\over\pi}\right)
\Gamma\left(1 + \lambda(4k-1) -{2\lambda u\over\pi}\right)}, \label{EqnI}
\end{eqnarray}
 \begin{eqnarray}
R_1(u,\xi)&=&{1\over\pi}\prod_{l=0}^{\infty}{{
\Gamma\left({1\over 2}+2l\lambda+{-\xi+u\lambda\over\pi}\right)
\Gamma\left({1\over 2}+2l\lambda+{\xi+u\lambda\over\pi}\right)
\over
\Gamma\left({1\over 2}+2l\lambda+\lambda+{-\xi+u\lambda\over\pi}\right)
\Gamma\left({1\over 2}+2l\lambda+\lambda+{\xi+u\lambda\over\pi}\right)
}}\times \nonumber \\
&\times &{\Gamma\left({1\over
2}+2l\lambda+\lambda+{\xi-u\lambda\over\pi}\right)
\Gamma\left({1\over 2}+2l\lambda+\lambda-{\xi+u\lambda\over\pi}\right)
\over
\Gamma\left({1\over 2}+2l\lambda+2\lambda+{\xi-u\lambda\over\pi}\right)
\Gamma\left({1\over 2}+2l\lambda+2\lambda-{\xi+u\lambda\over\pi}\right)}.
\label{EqnII}
\end{eqnarray}
The  poles of $P_{\pm}$ located in the physical domain $0<u<\pi/2$ at
$u_n=\pm{\xi\over\lambda}-{2n+1\over 2\lambda}\pi$
correspond to the ``boundary'' bound states of the theory. The latter exist in
the
soliton (resp. anti-soliton) scattering channel if $\xi>0$ (resp. $\xi<0$), and
their energy is $E_n=M_s\cos u_n$. Note that the
``physical'' values of the parameter $\xi$ are bounded \cite{SSW}:
$|\xi|<4\pi^2/\beta^2$.
In  the semi-classical limit
of the quantum field theory (\ref{lagr}), $\beta\to 0$,  the
principal (``tree'') approximation to the amplitudes
(\ref{reflamp}) has the following form \cite{SSW}:
\be
 P_{\pm}(\theta)=\exp{(\pm  i\xi + i|\xi|)}
{S(\theta;0)[S(2\theta;0)]^{1/2}\over
[S(\theta;\beta^2\xi/8\pi)S(\theta;-\beta^2\xi/8\pi)]^{1/2}}, \label{semiclass}
\ee
where
\be
S(\theta; y)=\exp{\left({8i\over\beta^2}\int_0^{\theta}
dv\ln\tanh^2{v+iy\over 2}
\right)},
\ee
$S(\theta;0)$ being the semi-classical approximation to the bulk
soliton-soliton
S-matrix \cite{ZZ}.

Our purpose here is
to show that the non-relativistic dynamics of quantum sine-Gordon solitons
in the presence of a boundary is described by the generalized Calogero-Moser
Hamiltonian:
\begin{eqnarray}
\hat{H} =~-~{1\over 2M_s} \sum_{i=1}^{N} {d^2\over dx_i^2}~-~{1\over 2M_s}
 \sum_{j=1}^{M} {d^2\over dy_j^2}~+~\sum_{i<i'}^{N}\left( V_{AA}\left(
 x_i-x_{i'}\right) +V_{AA}\left( x_i+x_i'\right)\right) \nonumber \\
+~\sum_{j<j'}^{M}\left( V_{AA}\left( y_j-y_{j'}\right) +V_{AA}\left( y_j+y_{j'}
\right)\right) ~+~\sum_{i=1}^{N}\sum_{j=1}^{M}\left( V_{A\bar{A}}\left( x_i-y_j
\right)+V_{A\bar{A}}\left( x_i+y_j\right)\right)
\nonumber \\
 +~\sum_{i=1}^{N} W_A\left( x_i\right) ~+~
\sum_{j=1}^{M} W_{\bar{A}}\left( y_j\right). \label{cmoser}
\end{eqnarray}
Here $V_{AA}$ and $V_{A\bar{A}}$ are bulk nonrelativistic potentials obtained
long ago in \cite{ZZ, Kor}
\be
V_{AA}\left( x\right)~=~{\alpha_0^2\over{M_s}}
{\rho(\rho -1)\over\sinh^2\alpha_0 x} ,\quad
V_{A\bar{A}}\left( x\right) ~=~-{\alpha_0^2\over{M_s}}{\rho(\rho-1)\over
\cosh^2\alpha_0 x},
\label{bulkV}
\ee
with
\be
\rho =  {8\pi \over{\beta^2}} , \label{idI}
\ee
\be
\alpha_0 = {m_0\over 2} , \label{idII}
\ee
 and $W_{A}$ and $W_{\bar{A}}$ are boundary potentials of the \ptp \cite{PT}
 type
\begin{eqnarray}
 W_A\left( x\right)&=&{\alpha_0^2\over{2M_s}}\left(
{\mu(\mu -1)\over\sinh^2\alpha_0 x} - {\nu(\nu-1)\over\cosh^2\alpha_0
x}\right),
 \nonumber \\
  W_{\bar{A}}\left( x\right)&=&{\alpha_0^2\over{2M_s}}\left({\nu(\nu
 -1)\over\sinh^2\alpha_0 x} -{\mu(\mu-1)\over\cosh^2\alpha_0 x}\right),\qquad
\mu >1, \nu >1 . \label{ptell}
\end{eqnarray}
We will show below that $\mu$ and $\nu$ are related to the parameter $\xi$
of the sine-Gordon model (\ref{lagr}) as follows:
\be
{\nu - \mu \over 2} = {\xi\over\pi} . \label{idIII}
\ee

Note that the translational invariance of the Hamiltonian (\ref{cmoser})
is broken not only
by the boundary potentials, but also by the interaction of particles with their
mirror images. This is very natural from the point of view of the underlying
sine-Gordon theory, since it can be shown \cite{SSW}, that the one soliton
 problem on a half-line is equivalent to the three-soliton bulk problem, with
one of the particles staying at $x=0$, and the other two being ``generalized
mirror images'' of each other (the B\"acklund transformation generalizing the
 method of images was constructed in \cite{Tar}). The analogy becomes exact if
 we take $\varphi_0=0$. Then the ordinary method of images works, and it is
 obvious that the system of $N+M$ solitons on a half-line is equivalent to the
 system of $2(N+M)$ solitons on a line with symmetric initial conditions. Hence
 the corresponding nonrelativistic Hamiltonian can be obtained from the known
\cite{ZZ,Kor} nonrelativistic bulk Hamiltonian. One can easily see that the
result is just (\ref{cmoser}) with $\mu=\nu\, ,\, \mu(\mu-1)=\rho(\rho-1)/4$.
Now let us turn to the case of arbitrary $\varphi_0.$

To establish the equivalence we will show that the $S$-matrices of the quantum
 sine-Gordon theory and the model (\ref{cmoser}) coincide in the appropriate
 limit. The system (\ref{cmoser}) is integrable both at the classical and
quantum levels \cite{IM,VanD}. To see this one takes the hyperbolic-type
 Calogero-Moser Hamiltonian for $N+M$
 particles based on the $BC_{N+M}$ root system \cite{PeOlsh}
 and shifts the  coordinates of the particles $N+1,...,N+M$ by $i\pi/ 2$.
The result is (\ref{cmoser}). Integrability means that the system admits a Lax
 representation and has $N+M$ integrals in involution. Moreover, since as
 $t\to\pm\infty$ these integrals reduce asymptotically to symmetric
 polynomials in particles' momenta, one can use the standard argument
\cite{Kul}
 to show that the $S$-matrix is factorized. A small modification arises due to
 the presence of the boundary; namely, one can consider the particles'
 collisions both very far from the boundary where the problem is reduced to the
 bulk one, and near the boundary where  the colliding particles have enough
time
to reflect and go to $x= +\infty$. In the first case factorization
gives the nonrelativistic Yang-Baxter equation for the bulk $S$-matrix
 \cite{ZZ}, while in the second case we get exactly the boundary Yang-Baxter
equation of \cite{GZ}, which in the Dirichlet case allows to express the
boundary $S$-matrix of the anti-kink through that of the kink \cite{GZ}. The
 unitarity requires the latter to be a pure phase, but otherwise leaves it
undetermined. So in both theories the $S$-matrix is factorized and fully
determined by the bulk two-particle $S$-matrix and the boundary $S$-matrix.
Thus to establish the equivalence it is sufficient to show that these
 $S$-matrices coincide when the nonrelativistic limit is taken.

Let us comment first on the properties of the one-particle Schr\"odinger
 equation with the \ptp potential $W_A(x)$.
The energy of the bound states, which appear when $\nu>\mu+1$, is given by
$E_n=-{\alpha_0^2\over 2M_s}(\nu-\mu-1-2n)^2$, where $n=0,1,2...$ For a fixed
value of $\nu-\mu$ there are in total $\left[{\nu-\mu-1\over 2}\right]$  bound
states. The reflection coefficient, which is a pure phase, can be obtained
to be equal to
\be
S_A(k)={
\Gamma\left({ik\over\alpha_0}\right)
\Gamma\left({1\over 2}+{\mu-\nu\over 2}-{ik\over 2\alpha_0}\right)
\Gamma\left({\mu+\nu\over 2}-{ik\over 2\alpha_0}\right)
\over
\Gamma\left(-{ik\over\alpha_0}\right)
\Gamma\left({1\over 2}+{\mu-\nu\over 2}+{ik\over 2\alpha_0}\right)
\Gamma\left({\mu+\nu\over 2}+{ik\over 2\alpha_0}\right) } \label{smatr}
\ee
This expression has ``physical'' poles  on
the upper imaginary half-axis
in the complex momentum plane which correspond to the bound states. Besides, it
has poles at the points $k_n=(1+n)\alpha_0$
that come from the first $\Gamma$-function in the
numerator of (\ref{smatr}). The latter set of poles is infinite and does not
correspond to any bound states of the theory. The $S$-matrix for the potential
$W_{\bar{A}}$ can be obtained from (\ref{smatr}) by the substitution $\mu
\leftrightarrow \nu$. One can see that $S_A$ and $S_{\bar{A}}$ satisfy indeed
the boundary Yang-Baxter equation of \cite{GZ}
\be
S_A \cos \left({\pi\over 2}(\nu-\mu)-\lambda u\right) = S_{\bar{A}}
 \cos\left({\pi\over 2}(\nu-\mu)+\lambda u\right).      \label{noref}
\ee

The nonrelativistic limit of (\ref{reflamp}) corresponds to the values
 $\theta\ll 1 .$ Simultaneously we must take the limit $\beta\ll 1,$ so that
 $M_s={{8m_0}\over\beta^2}\gg m_0$ (otherwise the $S$-matrix becomes $1$ and we
 do not get anything interesting.) Note that this is not a quasiclassical
limit, since the latter corresponds to ${k\over\alpha_0}\gg 1$, where
\be
k\equiv M_s\theta, \label{bydef}
\ee
whereas according to (\ref{idII}) ${k\over{\alpha_0}} ={
16\theta\over{\beta^2}}$, which is not necessarily large. We recover the
quasiclassical
limit in the region $\beta^2\ll\theta\ll 1$.  In what follows we assume that
$\xi$ is positive and $\beta\varphi_0\sim 1$, so that $\xi$ scales as
 $1/\beta^2$. Then $P_+$ has poles corresponding to the boundary bound states,
and the energy of the bound states lying close to the edge of the continuous
spectrum in the theory (\ref{lagr}) becomes
$E_n\simeq M_s - {m_0\pi\over 8\lambda}\left({2\xi\over\pi}-1-2n\right)^2$.
 $P_-$ does not have poles in the physical region. One can easily see then that
 (\ref{ptell}),(\ref{smatr}) describe correctly the spectrum of the boundary
 bound states provided  that equations (\ref{idII}),(\ref{idIII}) are
 fulfilled. To complete the identification of the boundary $S$-matrices we have
 to compare the phase shifts. Since $\xi$ scales as $1/\beta^2$, by virtue
 of (\ref{idIII}) the expression (\ref{smatr}) can be rewritten as
 $S_{NR}(k,\nu-\mu) f(\theta )$, where
\be
S_{NR}(k,\nu-\mu)={
\Gamma\left({ik\over\alpha_0}\right)
\Gamma\left({\mu-\nu\over 2}-{ik\over 2\alpha_0}\right)
\over
\Gamma\left(-{ik\over\alpha_0}\right)
\Gamma\left({\mu-\nu\over 2}+{ik\over 2\alpha_0}\right) } \label{slim}
\ee
is meromorphic and contains the poles located in the arbitrarily small
 neighbourhood of $\theta=0$ as $\beta\to 0$, whereas the factor $f(\theta)$
can
 be expanded into the power series $f(\theta)=1+\sum_{l=1}^{\infty}a_l\theta^l$
with all the coefficients and a radius of convergence $\sim 1$ as $\beta
\rightarrow 0$. In the same limit $P_{\pm}$ can be
factorized analogously: $P_{\pm}=S_{NR}(k,\pm 2\xi/\pi) f_{\pm}(\theta)$ with
$f_{\pm}$ admitting expansions of the form
 $f_{\pm}(\theta)=1+\sum_{l=1}^{\infty}a_l^{\pm}\theta^l$ with all the
coefficients and the radius  of convergence $\sim 1$ in the limit $\beta
\rightarrow 0$. Therefore the boundary $S$-matrices agree when
 $\theta\ll 1$, and $S_{NR}$ represents the nonrelativistic limit of the
boundary $S$-matrix of the sine-Gordon theory.
One can check that the same statements are true for the bulk two-particle
sine-Gordon $S$-matrix of \cite{ZZ} and the $S$-matrix of the particles
 interacting via the potentials (\ref{bulkV}), provided that (\ref{idI}),
(\ref{idII}) are satisfied. (This result was first established in \cite{Kor} in
the quasiclassical approximation and later confirmed in \cite{ZZ}. Note that
 our approach allows to give an exact sense to the statement that the
 nonrelativistic limit of the bulk sine-Gordon theory is the hyperbolic-type
Calogero-Moser model.) Thus the equivalence of (\ref{cmoser}) and the
nonrelativistic limit of (\ref{lagr}) is established.

It is instructive
to compare also the combined nonrelativistic/quasiclassical limit of
the S-matrix of  (\ref{lagr})
 with the \ptp S-matrix in the regime  $\beta^2\ll\theta\ll 1$
 (i.e. $k\gg m_0$) without applying to the exact formula  (\ref{reflamp}),
similar to how it was first done in \cite{Kor} for the bulk
 soliton scattering. This means that one should first take the limit
 $\beta\to 0$ and then, using (\ref{semiclass}), pass to the limit
$\theta\to 0$. Expanding the integrals in (\ref{semiclass}) and using the
 asymptotic formulas for the $\Gamma$-functions in (\ref{smatr}) we get
 for  (\ref{semiclass}) and (\ref{smatr}) in the principal order the
following result:
\be
P_+(k)=e^{2i\xi sign(k)+{4ik\over m_0}\ln{k\beta^2\over m_0}},\qquad
P_-(k)=e^{{4ik\over m_0}\ln{k\beta^2\over m_0}}, \label{asymp}
\ee
once again confirming the equivalence of the two theories.
Note that it is impossible to determine $\mu$ and $\nu$ separately, since
in our limit the boundary $S$-matrices in both theories depend only on the
difference $\mu-\nu$.

In conclusion we would like to mention that the identification of the
 nonrelativistic limit in the case of the most general integrable boundary
condition \cite{GZ} requires a nontrivial generalization of the
 Calogero-Moser Hamiltonians. Indeed, if in (\ref{lagr}) $M<\infty$, then the
 boundary $S$-matrix does not conserve the topological charge, and we are not
 aware of any integrable nonrelativistic model which allows such a process.

One of the authors (S.S.) would like to thank H.Saleur, S.N.M. Ruijsenaars and
 V.Korepin for helpful discussions. The work of A.K. was supported in part by
 the U.S. Dept. of Energy under Grant No. DE-FG03-92-ER40701, while the
work of S.S. was supported by the grant NSF-PHY-9357207.

 \end{document}